# A Text Steganography Method Using Pangram and Image Mediums

Youssef Bassil

LACSC – Lebanese Association for Computational Sciences
Registered under No. 957, 2011, Beirut, Lebanon

youssef.bassil@lacsc.org

**Abstract**—Steganography is the art and science of writing hidden messages in such a way that no one apart from the sender and the receiver would realize that a secret communicating is taking place. Unlike cryptography which only scrambles secret data keeping them overt, steganography covers secret data into medium files such as image files and transmits them in total secrecy avoiding drawing eavesdroppers' suspicions. However, considering that the public channel is monitored by eavesdroppers, it is vulnerable to stego-attacks which refer to randomly trying to break the medium file and recover the secret data out of it. That is often true because steganalysts assume that the secret data are encoded into a single medium file and not into multiple ones that complement each other. This paper proposes a text steganography method for hiding secret textual data using two mediums; a Pangram sentence containing all the characters of the alphabet, and an uncompressed image file. The algorithm tries to search for every character of the secret message into the Pangram text. The search starts from a random index called seed and ends up on the index of the first occurrence of the character being searched for. As a result, two indexes are obtained, the seed and the offset indexes. Together they are embedded into the three LSBs of the color channels of the image medium. Ultimately, both mediums mainly the Pangram and the image are sent to the receiver. The advantage of the proposed method is that it makes the covert data hard to be recovered by unauthorized parties as it uses two mediums, instead of one, to deliver the secret data. Experiments conducted, illustrated an example that explained how to encode and decode a secret text message using the Pangram and the image mediums. As future work, other formats of files for the second medium are to be supported enabling the proposed method to be generically employed for a wide range of applications.

—————————— ◆ ——————————

## 1 INTRODUCTION

TODAY, the Internet is a key communication infrastructure for inter-connecting individuals across the globe, allowing them to send, receive, and share data among each other. As this trend developed into an everyday activity, securing sensitive data became of matter of great concern [1]. Cryptography took the IT industry by storm when first applied to secure Internet traffic. One of the most common encryption protocols for the Internet is the SSL (Secure Socket Layer) which encrypts the network traffic at the Application Layer using asymmetric cryptography [2]. In effect, SSL allows safe data transmission between browsers and online applications while impeding packet eavesdropping, data tampering, and other malicious attacks. Basically, cryptography provides data confidentiality by scrambling the content of data turning it into a new form that is incomprehensible by unauthorized parties [3]. Despite the fact that eavesdroppers cannot understand the encrypted data being transmitted, they are very aware that a secret communication is taking place. As a result, steganography started to attract the attention of computer researchers and users. In fact, the goal of steganography is to hide the very existence of communication by embedding the secret data to transmit into a digital media file called carrier file such as an image or text file [4]. The carrier file is the only visible object to be communicated between the sender and the receiver. Since the secret data are invisible and covert into the carrier file, no one apart from the concerned communicating parties is aware that a secret communication is happening [5]. The strength of steganography resides in how difficult the covert data can be detected and recovered by unauthorized parties. Steganalysts are specialists who are hired to identify suspicious files and detect whether or not they contain secret data, and if possible, recover this data. However, steganalysts often assume that the secret data are hidden into a single carrier file and not in multiple ones that complement each other. The reason for this is that the majority of the current steganography techniques follow the principle of using one carrier file to transmit secret data [6].

This paper proposes a new steganography method for text hiding. It pivots around using two user-defined mediums to encode and decode the secret data. The first medium is a Pangram sentence made up of English words; while, the second medium is an uncompressed image file. The proposed method is driven by an algorithm that works as follows: Every character in the input secret message is encoded using two indexes; a seed index that points to a random character in the Pangram sentence, and an offset index that denotes the distance between the seed index and the first occurrence of the character being encoded in the Pangram sentence. The maximum allowable length of the Pangram is 512 characters; hence, only 9 bits are required to represent the seed and the offset indexes. Both indexes are then embedded into the three LSBs of the color channels of the image medium. Ultimately, both mediums the Pangram and the image are sent to the receiver. The advantage of the proposed method is that it employs two mediums, instead of one, that complement each other to deliver the secret data, making the covert data so robust against stego-attacks.

## 2 FUNDAMENTALS OF STEGANOGRAPHY

Fundamentally, steganography refers to "secret writing" in Greek, and is the art and science of hiding information inside innocuous files such as images, audio files, and video files, in ways that avoid the detection of the hidden information [7]. The outcome of steganography is a covert channel of communication through which secret data can be transmitted in total secrecy avoiding drawing eavesdroppers' suspicions.

As an algorithmic model, steganography can be thought as the "Prisoners' Problem" [8]. In this model, two prisoners put in jail, Alice and Bob, want to communicate about an escape plan. The challenge is that they can only communicate through the warden of the prison, Wendy, who prohibits both Alice and Bob to communicate in code using standard cryptography. As a result, Alice and Bob devise a new method for secret communication called steganography. It is about hiding the message that needs to be communicated in an innocent-looking image. This image is then handed in by Bob to the warden Wendy to pass it along to Alice. Wendy, looking at the image, would not notice anything suspicious and would subsequently pass it along, not knowing that the pixels of the image encode the secret message. Alice, after receiving the image, would recover the secret message of Bob as she knows how the trick works. Formally, the steganography model can be mathematically defined as follows [9]: the original file into which the secret message is to be concealed is denoted by A, the secret message to hide by M, and the carrier file by C. Actually, the carrier file C is visually or audibly identical to the original file A but with one difference is that it has the message M embedded inside it. The steganography encoding algorithm which is used to cover a secret message M into a file A, is denoted by S(A, M, Enc)=C where "Enc" denotes the encoding mode. On the other hand, the steganography decoding algorithm which is used to recover the secret message M out of the carrier file C, is denoted by S(C, Dec)=M where "Dec" denotes the decoding mode. Figure 1 depicts the basic mathematical model of steganography.

## 3 STATE-OF-THE-ART IN TEXT STEGANOGRAPHY

Hiding information in plain text can be done in many different ways. Some techniques consist of changing the outline of the carrier text such as adding whitespaces or altering the case of certain characters so as to represent secret text [10]. Others, consist of relating the characters to hide with the characters of the carrier text, creating a reference dictionary that maps words from the secret text with words from the carrier text [11]. This section sheds the light on the various techniques used in text steganography including hiding by selection [12], hiding in HTML [13], line and word shifting [14], hiding using whitespace [15], semantic-based hiding [16], and abbreviation-based hiding [17] techniques.

**Hiding by Selection**: The selection technique selects certain characters in the carrier text to convey the characters of the secret message such as selecting the first character of every word in the carrier text or the second character of every other word. Now, in order to recover the concealed secret message, all first characters of the words of the carrier text are extracted and concatenated together, producing the exact original message. A variation of this technique can be performed by selecting the first character from the first word, the second character from the second word, the third character from the third word, and so forth, until the characters of the message to hide are exhausted. NULL cipher [18] is in essence based on the selection technique as it constructs an unsuspicious plaintext having the secret message as part of its characters. For example, in World War I, the German embassy in United States sent a telegraph to Berlin stating "**P**resident's **E**mbargo **R**uling **S**hould **H**ave **I**mmediate **N**otice. **G**rave **S**ituation **A**ffecting **I**nternational **L**aw. **S**tatement **F**oreshadows **R**uin **O**f **M**any **N**eutrals. **Y**ellow **J**ournals **U**nifying **N**ational **E**xcitement **I**mmensely". It is by reading the first character of every word of this statement, that the secret message can be revealed as "Pershing Sails from NY June I" [19]. The drawback of this method is that it requires a huge volume of text to hide a small message of few words.

**HTML Documents**: Secret text can be easily concealed within HTML documents because HTML tags are case insensitive. For instance, the tags <a title="clients">, <a TITLE="clients">, and <a TitlE ="Center">, are all the same and have the same effect on the rendering of the document. Text steganography applied in HTML documents can be performed by changing the case of the letters that make up the HTML tags. In particular, the secret message is represented by the capital version of the tags' letters, or vice versa depending on the algorithm being used. As for the recovering process, all the capital letters from the HTML document have to be captured and concatenated together in order to produce the original covert message.

**Line and Word Shifting**: In this technique, text lines are shifted vertically and words are shifted horizontally by a fixed space of $n$ inches. That way, the distance between lines and words would convey the hidden characters. For instance, letter A can be encoded as a 0.01 inch space between two text lines. Similarly, letter B can be encoded as a 0.02 inch space between two other text lines. In effect, this technique is more suitable for printed text than for digital text, since printed spaces can be physically measured unlike their digital counterparts.

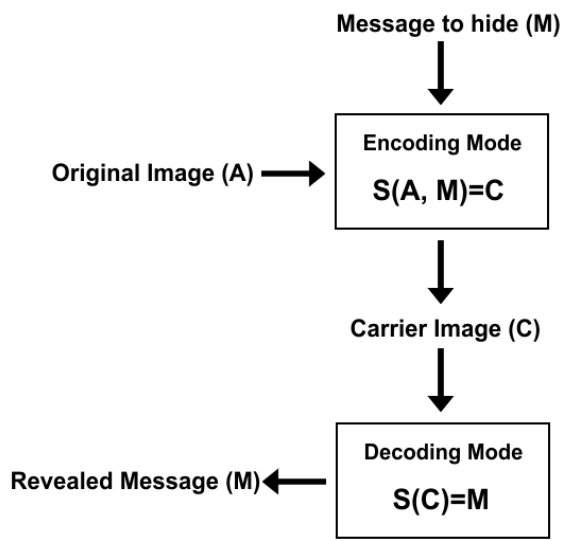

Fig. 1. Mathematical Model of Steganography

**Hiding using Whitespace**: Its concept is very straightforward. A message to hide is first converted into a binary format. Then, every bit whose value is 1 is represented by an extra whitespace between a particular set of two words in the carrier text; whereas, every bit whose value is 0 leaves the original single whitespace between the next particular set of two words. For example, "the boy went to school today " can be deciphered as "101001". In fact, two spaces exist between "the" and "boy", between "went" and "to", and between "today" and the end of the sentence. This results in a bit of value 1 in positions 0, 2, and 5 respectively. In contrast, only a single space exists between "boy" and "went", between "to" and "school", and between "school" and "today". This results in a bit of value 0 in positions 1, 3, and 4 respectively. Basically, the whitespace technique is very suspicious as a normal reader would right away notice the existence of some extra whitespaces in the text. Additionally, this method cannot encode too much information especially in small text.

**Semantic-Based Hiding:** The semantic-based text steganography technique uses synonyms of words to hide the secret information in the carrier text. For instance, the secret message "the boy went to school today" can be encoded using the semantic approach as "the child went to college today".

**Abbreviation-Based Hiding:** This technique uses a lexical dictionary containing words along with their abbreviations. These abbreviations are either labeled 0 or 1. While performing steganography, if a word in the carrier text is found in the dictionary, it is substituted by its abbreviation based on the current bit to hide. Different values of bits have different corresponding abbreviations. Some examples of these abbreviations can be "ASAP" for "As Soon As Possible", "CU" for "see you", "gr8" for "Great" etc. A variation of this method is the one proposed by [10], which consists of changing the spelling of words based on their American and British spellings. For example, "Favorite" is designated by 1 while "Favourite" is designated by 0, and "Center" is designated by 1 while "Centre" is designated by 0.

## 4 PROPOSED SOLUTION

This paper proposes a new steganography method for text hiding. It revolves around using two mediums to transmit secret textual data from sender to receiver. The first medium is a user-defined Pangram English text sentence made up of a maximum of 512 characters including letters, digits, and special characters. The second medium is a digital uncompressed image file mainly a BMP type image file. The proposed algorithm works as follows: Every character in the secret message is encoded using two indexes; a seed index that points to a random character in the Pangram sentence (the first medium), and an offset index that refers to the distance between the seed index and character in the Pangram sentence that is equal to the one being encoded. As by design, the Pangram is user-defined and has a maximum length of 512 characters; thus, only 9 bits are required to create the indexes. Both indexes are then embedded into a carrier image (the second medium) such as the seed index into a given pixel; while, the offset index into the very next pixel. The exact bits where the seed and the offset indexes are stored in the three LSBs of the color channels of the carrier image, i.e., 3 bits in the Red channel, 3 bits in the Green channel, and 3 bits in Blue channel per every pixel. Finally, both mediums, the Pangram and the image are sent to the receiver.

### 4.1 The First Medium – The Pangram Sentence

The proposed method uses as a first medium to deliver secret data, a correctly structured Pangram English sentence made up of English words. A Pangram, which means "every letter" in Greek, is a sentence composed of every letter of the alphabet at least once [20]. Pangrams have been used to test telecommunication devices such as in telegraphy, to display font samples such as the Microsoft font viewer, among other applications. One of the most widely known English Pangram is "The quick brown fox jumps over the lazy dog" which contains all the letters of the English alphabet. The proposed steganography method requires its Pangram to have not only all possible English letters, but also digits and special characters so that it can match any possible character from the secret message to hide. The maximum length of the Pangram is 512 characters with the possibility of having duplicate characters. Formally, every character $s$ in the Pangram has an index denoted by $i$ which uniquely points to a particular character in the Pangram itself. The Pangram is represented as PAN=$\{s_{i=0}, s_{i=1}, s_{i=2}, s_{i=3},…,s_{i=511}\}$. For instance, the Pangram "The quick brown fox jumps over the lazy dog" can be represented as PAN=$\{$ T,h,e,q,u,i,c,k,b,r,o,w,n,f,o,x,j,u,m,p,s, o,v,e,r, t,h,e, l,a,z,y, d,o,g$\}$. In effect, the Pangram is used to match a character from the secret message with an equal one from the Pangram. The matching process uses a linear search algorithm starting from a random seed index denoted by SEED till an offset index denoted by OFFSET representing the distance from the SEED index to the actual position of the matching character in PAN. As the maximum length of the Pangram is 512 characters, the indexes are each 9 bits long ($2^9$=512). Post to the completion of the encoding process, the Pangram is sent to the receiver along with the second medium namely the carrier image.

### 4.2 The Second Medium – The Carrier Image

The second medium is a digital uncompressed image file of type BMP. It is a 24-bit colored image whose pixels are each made up of three color channels: Red, Green, and Blue channels (RGB). This medium is to act as a carrier file to embed the SEED and the OFFSET indexes generated out of the first medium that is the Pangram sentence. The SEED is embedded into a particular pixel; while, the OFFSET in the very next one. The bit locations into which these indexes are to be stored is the three LSBs (Least Significant Bits) of the color channels that constitute the pixels of the carrier image. The hiding capacity is around 2.08% characters of the total carrier image size (i.e., two 9 bits out of two 24-bit pixels are required since every two pixels encode a single character, one pixel to store the SEED and one to store the OFFSET). For instance, a 1MB BMP image can encode 174762 characters ((1024KB*1024B*8b)/48). Formally, the second medium can be represented as IMG=$\{p_0[R_0, G_0, B_0]$ , $p_1[R_1, G_1, B_1]$, $p_2[R_2, G_2, B_2]$, $p_{n-1}[R_{n-1}, G_{n-1}, B_{n-1}]$ $\}$ where $p_i$ is the $i^{th}$ pixel in the carrier image IMG, $R_i$, $G_i$, and $B_i$ are the three color channels that compose pixel $p_i$, and $n$ is the total number of pixels in IMG. Every $p_i$ is composed of 24 bits organized into three parts each of which is 8 bits and represents a particular color channel. Pixel $p_i$ is denoted as $p_i = \{ R_i [r_0,r_1,r_2,r_3,r_4,r_5,r_6,r_7], G_i [g_0,g_1,g_2,g_3,g_4,g_5,g_6,g_7], B_i [b_0,b_1,b_2,b_3,b_4,b_5,b_6,b_7] \}$ where $i$ is the index of the $i^{th}$ pixel, $r_j$ is the $j^{th}$ bit of channel R, $g_j$ is the $j^{th}$ bit of channel G, and $b_j$ is the $j^{th}$ bit of channel B. Eventually, when covering the SEED index in IMG, its bits would replace $r_5, r_6, r_7$ in $R_i$, $g_5, g_6, g_7$ in $G_i$, and $b_5, b_6, b_7$ in $B_i$, and the bits of the OFFSET index would replace $r_5, r_6, r_7$ in $R_{i+1}$, $g_5, g_6, g_7$ in $G_{i+1}$, and $b_5, b_6, b_7$ in $B_{i+1}$.

### 4.3 The Proposed Algorithm

The proposed algorithm involves several steps to be completed in order to conceal an input text message M using two user-defined mediums, a Pangram sentence denoted by PAN and an image file denoted by IMG. These steps are executed one after the other in the following order:

1. An input secret text message denoted by $M=\{m_0,m_1,m_2,...m_{n-1}\}$ where $m$ is single character in M and $n$ is the total number of characters in M, is fed to the algorithm.
2. A random number is chosen called seed index denoted by SEED such as SEED<512. Its purpose is to point to random character $s$ in PAN which is denoted by $PAN=\{s_{i=0}, s_{i=1}, s_{i=2}, s_{i=3},...,s_{i=511}\}$ where $s$ is a single character in PAN, $i$ is an index pointing to the $i^{th}$ character $s$, and 512 is the maximum number of characters that a PAN can have.
3. Starting from the SEED index, PAN is searched linearly for the first occurrence of a particular character $m$ from the input secret message M. Once found, an offset index denoted by OFFSET is calculated as the distance from the SEED index to the index of $m$ in PAN such as OFFSET=index_of_m-SEED. As 512 characters is the maximum length of PAN, 9 bits are required to represent the indexes SEED and OFFSET. It is worth noting that PAN is regarded as circular, which means that last character comes next to the first character. As a result, SEED can be greater than *index_of_m*; thus yielding to a negative result for OFFSET. However, as the OFFSET must always be a positive number, the maximum length of PAN is added to the obtained results, such as IF index_of_m-SEED<0 THEN OFFSET=index_of_m-SEED+*max_length_pan*.
4. The indexes SEED and OFFSET are converted into binary yielding to 9 bits for each, such as $SEED=\{e_0,e_1,e_2,e_3,e_4,e_5,e_6,e_7,e_8\}$ where $e$ is a single bit in SEED, and $OFFSET=\{k_0,k_1,k_2,k_3,k_4,k_5,k_6,k_7,k_8\}$ where $k$ is a single bit in OFFSET.
5. Working with the second medium namely the carrier image, it is denoted by $IMG=\{p_0[R_0, G_0, B_0], p_1[R_1, G_1, B_1], p_2[R_2, G_2, B_2], p_{n-1}[R_{n-1}, G_{n-1}, B_{n-1}]\}$ where $p_i$ is the $i^{th}$ pixel in the carrier image IMG, $R_i$, $G_i$, and $B_i$ are the three color channels that compose pixel $p_i$, and $n$ is the total number of pixels in IMG. Every $p_i$ is composed of 24 bits organized into three parts each of which is 8 bits and represents a particular color channel. $p_i$ is denoted as $p_i = \{ R_i [r_0,r_1,r_2,r_3,r_4,r_5,r_6,r_7], G_i [g_0,g_1,g_2,g_3,g_4,g_5,g_6,g_7], B_i [b_0,b_1,b_2,b_3,b_4,b_5,b_6,b_7] \}$ where $i$ is the index of the $i^{th}$ pixel, $r_j$ is the $j^{th}$ bit of component R, $g_j$ is the $j^{th}$ bit of component G, and $b_j$ is the $j^{th}$ bit of component B.
6. The random SEED index is stored in the three LSBs of the three color channels of pixel $p_t$; while, the OFFSET index is stored in the three LSBs of the three color channels of pixel $p_{t+1}$. The process is carried out such as $p_t=\{ R_t [r_0,r_1,r_2,r_3,r_4, e_0,e_1,e_2], G_t [g_0,g_1,g_2,g_3,g_4, e_3,e_4,e_5], B_t [b_0,b_1,b_2,b_3,b_4, e_6,e_7,e_8] \}$ and $p_{t+1}=\{ R_{t+1} [r_0,r_1,r_2,r_3,r_4, k_0,k_1,k_2], G_{t+1} [g_0,g_1,g_2,g_3,g_4, k_3,k_4,k_5], B_{t+1} [b_0,b_1,b_2,b_3,b_4, k_6,k_7,k_8] \}$
7. Step 2-6 are repeated until all characters in M are exhausted. Then, both PAN (Pangram sentence) and IMG (the carrier image) are sent to the receiver.

In order to recover the secret message M when both mediums are received by the receiver, the three LSBs of every pixel in the second medium (carrier image) are to be extracted in such a way that every extracted two pairs of 9 bits are grouped together as they represent the SEED and the OFFSET indexes. Afterwards, the second medium (Pangram sentence) is used to locate the secret characters using *index_of_m*=SEED+OFFSET. The character in the Pangram that is pointed to by *index_of_m* is the actual secret character. Doing so for all extracted SEED and OFFSET indexes would eventually yield to all secret characters being revealed.

## 5 THE PROPOSED METHOD AS A NEW MODEL FOR STEGANOGRAPHY

As previously discussed, a steganography model can be mathematically defined as having two processes: An encoding process carried out such as S(A, M, Enc)=C where A is the original medium into which the secret message M is to be concealed which would result in a modified medium called carrier medium denoted by C. And a decoding process carried out such as S(C, Dec)=M where M is the secret message M recovered out of the carrier medium C. Notwithstanding, the proposed method imposes a new mathematical model defined as: An encoding process carried out such as S(A, P, M, Enc)={C, P} where A is the first medium and P is the second medium, that both complement each other. And a decoding process carried out such as S(C, P, Dec)=M where M is the secret message recovered out using both mediums C and P. Figure 2 depicts the mathematical model for the proposed method.

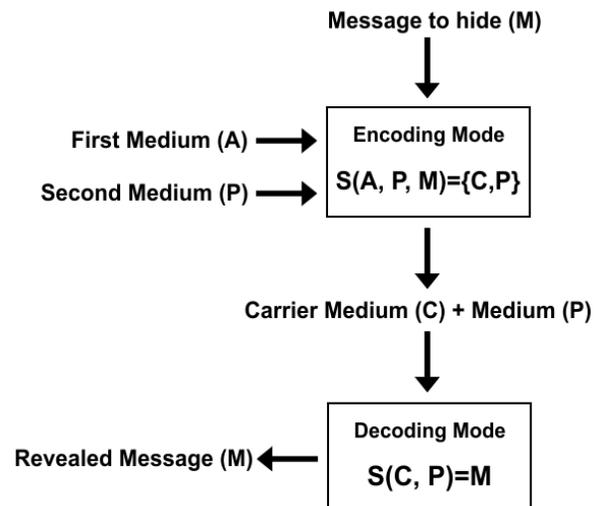

Fig. 2. Mathematical Model of the Proposed Method

## 6 EXPERIMENTS & RESULTS

In this section, an example is illustrated about how to hide a secret message M="KILL JOE" using a Pangram sentence and a carrier image. The carrier image is a 24-bit BMP image having a resolution of 800x600. The Pangram is equals to PAN={*The apple is the pomaceous fruit which requires special care and there are more than 7,500 known cultivars of apples. Alexander the Great is credited with finding dwarfed apples in Kazakhstan. The United States is just the second producer, with more than 6% of world production; around $6 billion. The apple forms a tree, reaching 3 to 12 meters (9.8 to 39 ft) tall, with a broad, often densely twiggy crown. The leaves are arranged 1.2 to 2.4 in broad on a 0.79 to 2.0 in petiole with just an acute tip*}. Obviously, PAN contains all the letters of the English alphabet, in addition to all digits of the decimal system and some special characters. As the Pangram is user-defined, a user can craft his own Pangram depending the characters needed, though the above mentioned Pangram can be enough for most applications. The length of PAN is 504 characters (from index 0 to 503) including spaces so it is below 512 (the maximum allowed length); and thus, only 9 bits are required to represent any position. Following, are the results obtained.

1. The message to hide M is denoted by M={K, I, L, L, SPACE, J, O, E}
2. For every character in M, a random SEED index is selected to point to a random character in PAN such as SEED<504. Then PAN is searched for the first occurrence of a particular character in M. The index of the first occurrence is denoted by *index_of_m*. Finally, the offset (distance) between the SEED and *index_of_m* is calculated as: IF *index_of_m*-SEED<0 THEN OFFSET=*index_of_m*-SEED+*max_length_pan* ELSE OFFSET=*index_of_m*-SEED. Below are the results obtained for the OFFSETs:
   K → SEED=12, *index_of_K* =91, OFFSET=91-12=79
   I → SEED=1, *index_of_I*=10, OFFSET =10-1=9
   L → SEED=130, *index_of_L*=174, OFFSET =174-130=44
   L → SEED=340, *index_of_L*=363, OFFSET =363-340=23
   SPACE → SEED=50, *index_of_SPACE* =55, OFFSET =55-50=5
   J → SEED=2, *index_of_J*=214, OFFSET =214-2=212
   O → SEED=62, *index_of_O*=76, OFFSET =76-62=14
   E → SEED=500, *index_of_E*=2, OFFSET =2-500=-498+504=6
3. Every SEED and OFFSET are converted into a binary form of 9 bits.
   K → SEED=000001100, OFFSET=001001111
   I → SEED=000000001, OFFSET =000001001
   L → SEED= 010000010, OFFSET = 000101100
   L → SEED= 101010100, OFFSET = 000010111
   SPACE → SEED= 000110010, OFFSET =000000101
   J → SEED=000000010, OFFSET = 011010100
   O → SEED= 000111110, OFFSET =000001110
   E → SEED= 111110100, OFFSET =000000110
4. The SEED is stored in the three LSBs of the color channels of pixel $p_t$; while, the OFFSET is stored in the three LSBs of the color channels of pixel $p_{t+1}$. Since there are 8 characters in M, 16 consecutive pixels are required (8 for the SEEDs and 8 for the OFFSETs).

Following are the results obtained. r, g, and b are the original bits of the carrier image unaltered:
$p_0$={ $R_0$ [$r_0,r_1,r_2,r_3,r_4$, 0,0,0], $G_0$ [$g_0,g_1,g_2,g_3,g_4$, 0,0,1], $B_0$ [$b_0,b_1,b_2,b_3,b_4$, 1,0,0] }
$p_1$={ $R_1$ [$r_0,r_1,r_2,r_3,r_4$, 0,0,1], $G_1$ [$g_0,g_1,g_2,g_3,g_4$, 0,0,1], $B_1$ [$b_0,b_1,b_2,b_3,b_4$, 1,1,1] }
$p_2$={ $R_2$ [$r_0,r_1,r_2,r_3,r_4$, 0,0,0], $G_2$ [$g_0,g_1,g_2,g_3,g_4$, 0,0,0], $B_2$ [$b_0,b_1,b_2,b_3,b_4$, 0,0,1] }
$p_3$={ $R_3$ [$r_0,r_1,r_2,r_3,r_4$, 0,0,0], $G_3$ [$g_0,g_1,g_2,g_3,g_4$, 0,0,1], $B_3$ [$b_0,b_1,b_2,b_3,b_4$, 0,0,1] }
$p_4$={ $R_4$ [$r_0,r_1,r_2,r_3,r_4$, 0,1,0], $G_4$ [$g_0,g_1,g_2,g_3,g_4$, 0,0,0], $B_4$ [$b_0,b_1,b_2,b_3,b_4$, 0,1,0] }
$p_5$={ $R_5$ [$r_0,r_1,r_2,r_3,r_4$, 0,0,0], $G_5$ [$g_0,g_1,g_2,g_3,g_4$, 1,0,1], $B_5$ [$b_0,b_1,b_2,b_3,b_4$, 1,0,0] }
$p_6$={ $R_6$ [$r_0,r_1,r_2,r_3,r_4$, 1,0,1], $G_6$ [$g_0,g_1,g_2,g_3,g_4$, 0,1,0], $B_6$ [$b_0,b_1,b_2,b_3,b_4$, 1,0,0] }
$p_7$={ $R_7$ [$r_0,r_1,r_2,r_3,r_4$, 0,0,0], $G_7$ [$g_0,g_1,g_2,g_3,g_4$, 0,1,0], $B_7$ [$b_0,b_1,b_2,b_3,b_4$, 1,1,1] }
$p_8$={ $R_8$ [$r_0,r_1,r_2,r_3,r_4$, 0,0,0], $G_8$ [$g_0,g_1,g_2,g_3,g_4$, 1,1,0], $B_8$ [$b_0,b_1,b_2,b_3,b_4$, 0,1,0] }
$p_9$={ $R_9$ [$r_0,r_1,r_2,r_3,r_4$, 0,0,0], $G_9$ [$g_0,g_1,g_2,g_3,g_4$, 0,0,0], $B_9$ [$b_0,b_1,b_2,b_3,b_4$, 1,0,1] }
$p_{10}$={ $R_{10}$ [$r_0,r_1,r_2,r_3,r_4$, 0,0,0], $G_{10}$ [$g_0,g_1,g_2,g_3,g_4$, 0,0,0], $B_{10}$ [$b_0,b_1,b_2,b_3,b_4$, 0,1,0] }
$p_{11}$={ $R_{11}$ [$r_0,r_1,r_2,r_3,r_4$, 0,1,1], $G_{11}$ [$g_0,g_1,g_2,g_3,g_4$, 0,1,0], $B_{11}$ [$b_0,b_1,b_2,b_3,b_4$, 1,0,0] }
$p_{12}$={ $R_{12}$ [$r_0,r_1,r_2,r_3,r_4$, 0,0,0], $G_{12}$ [$g_0,g_1,g_2,g_3,g_4$, 1,1,1], $B_{12}$ [$b_0,b_1,b_2,b_3,b_4$, 1,1,0] }
$p_{13}$={ $R_{13}$ [$r_0,r_1,r_2,r_3,r_4$, 0,0,0], $G_{13}$ [$g_0,g_1,g_2,g_3,g_4$, 0,0,1], $B_{13}$ [$b_0,b_1,b_2,b_3,b_4$, 1,1,0] }
$p_{14}$={ $R_{14}$ [$r_0,r_1,r_2,r_3,r_4$, 1,1,1], $G_{14}$ [$g_0,g_1,g_2,g_3,g_4$, 1,1,0], $B_{14}$ [$b_0,b_1,b_2,b_3,b_4$, 1,0,0] }
$p_{15}$={ $R_{15}$ [$r_0,r_1,r_2,r_3,r_4$, 0,0,0], $G_{15}$ [$g_0,g_1,g_2,g_3,g_4$, 0,0,0], $B_{15}$ [$b_0,b_1,b_2,b_3,b_4$, 1,1,0] }

5. Finally, both mediums the Pangram PAN and the modified carrier image IMG and are sent to the receiver.

Figure 3 depicts the carrier image medium before and after the hiding process.

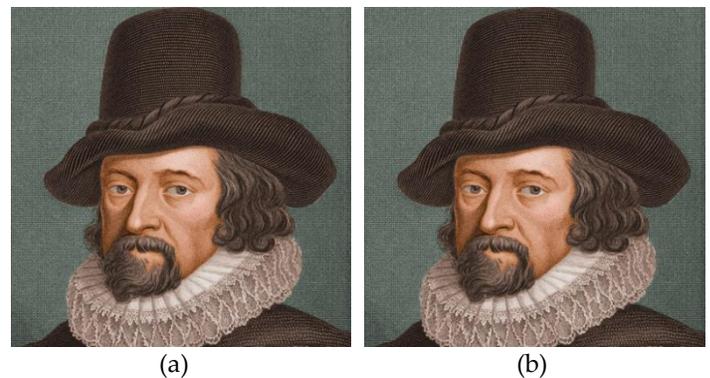

(a)          (b)
Fig. 3. Mediums before (a) and after (b) the hiding process

## 7 CONCLUSIONS & FUTURE WORK

This paper proposed a novel steganography method for textual data. It supports the covering and uncovering of plain text messages using two user-defined mediums: A Pangram sentence and an uncompressed image file. The image medium is used to embed the actual positions of the secret characters in the Pangram medium. The pro of this method is that it is resistant to stego-attacks as steganalysts often assume that the secret data are encoded using one medium and not using two mediums that complement each other. As a result, the sender can first send one of the mediums, and then later on, send the other one. All in all, it makes it difficult for eavesdroppers to infer how the system works and how to detect and recover the covert secret data.

As future work, other formats for the second medium are to be investigated in an attempt to enable the proposed method to be compatible with not only BMP files but also with JPG, GIF, WAV, MP4, among other types of files.

## ACKNOWLEDGMENT

This research was funded by the Lebanese Association for Computational Sciences (LACSC), Beirut, Lebanon, under the "Stealthy Steganography Research Project – SSRP2012".